\documentclass[aps,prl,twocolumn,amsmath,amssymb,superscriptaddress,showpacs,longbibliography]{revtex4-1}
\usepackage{graphicx}
\usepackage{dcolumn}
\usepackage{bm}

\usepackage{natbib}
\usepackage[
text={7in,9.5in},centering,
]{geometry}

\usepackage{epsfig}
\usepackage{amsmath}
\usepackage{amssymb}
\usepackage{textcomp}

\begin{document}

\title{Mapping the Structure of Directed Networks: Beyond the ``Bow-Tie'' Diagram}

\author{G. Tim\'ar}
 \email{gtimar@ua.pt}
 \affiliation{Departamento de F\'\i sica, Universidade de Aveiro and I3N, Campus Universit\'ario de Santiago, 3810-193 Aveiro, Portugal}

\author{A. V. Goltsev}
\email{goltsev@ua.pt}
 \affiliation{Departamento de F\'\i sica, Universidade de Aveiro and I3N, Campus Universit\'ario de Santiago, 3810-193 Aveiro, Portugal}
 \affiliation{A. F. Ioffe Physico-Technical Institute, 194021 St. Petersburg, Russia}

\author{S. N. Dorogovtsev}
 \affiliation{Departamento de F\'\i sica, Universidade de Aveiro and I3N, Campus Universit\'ario de Santiago, 3810-193 Aveiro, Portugal}
 \affiliation{A. F. Ioffe Physico-Technical Institute, 194021 St. Petersburg, Russia}

\author{J. F. F. Mendes}
 \affiliation{Departamento de F\'\i sica, Universidade de Aveiro and I3N, Campus Universit\'ario de Santiago, 3810-193 Aveiro, Portugal}


\begin{abstract}
We reveal a hierarchical, multilayer organization of finite components -- i.e., tendrils and tubes -- around the giant connected components in directed networks and propose efficient algorithms allowing one to uncover the entire organization of key real-world directed networks, such as the World Wide Web, the neural network of \emph{Caenorhabditis elegans}, and others.
With increasing damage, the giant components decrease in size while the number and size of tendril layers increase, enhancing the susceptibility of the networks to damage.
\end{abstract}


\maketitle

\begin{figure}[htpb!]
\centering
\includegraphics[width=6cm,angle=0.]{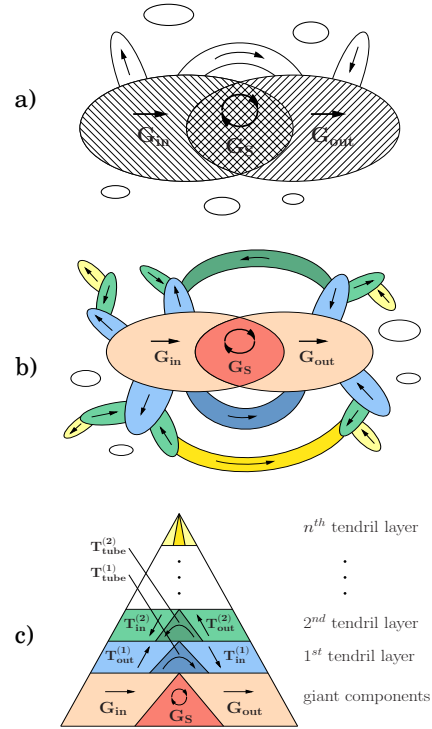}
\caption{ (a) The bow-tie diagram \cite{broder2000graph} for a directed graph with a giant strongly connected component ($G_S$), giant in-component and out-component ($G_{in}$ and $G_{out}$), and finite directed components (tendrils and tubes). Disconnected finite clusters are shown as open ovals.
(b) Schematic view of the complete structure of a directed
network. Different tendril layers are shown by different colors.
In general, there can be any number of tendril layers; we show three layers.
(c) The hierarchical structure of tendril layers.}
\label{fig:picture}
\end{figure}

Many real-world networks can be represented by directed graphs,
where each edge connecting two vertices is assigned one of two possible directions, or both.
Well-known examples are the World Wide Web (WWW), neuronal and metabolic networks, and many other systems \cite{newman2003structure,dorogovtsev2002evolution}.
Accounting for the link directedness is pivotal for understanding the structure and function of
such complex networks.
Directed networks have  certain structural properties in common
\cite{broder2000graph,newman2001random,dorogovtsev2001giant,schwartz2002percolation,boguna2005generalized}.
Any large directed network can be partitioned into several qualitatively different subgraphs:
(i) a giant strongly connected component and giant in-component and out-component, (ii) finite directed components (tendrils and tubes), and (iii) disconnected finite components.
Taking the union of the giant components, tendrils, and tubes and neglecting the edge directedness, we obtain the giant connected component of the undirected version of the graph under consideration.
Broder \emph{et al.} \cite{broder2000graph} represented the giant components of the WWW by use of the bow-tie diagram in Fig. \ref{fig:picture}(a), which is valid for an arbitrary directed graph.
Up to now, research of directed networks has been focused
mainly on the giant components, and has not touched tendril organization \cite{broder2000graph,newman2001random,dorogovtsev2001giant,boguna2005generalized}. However,
in sparse directed networks the total number of nodes in tendrils is a finite fraction of all nodes \cite{broder2000graph,dorogovtsev2001giant}.
One cannot fully understand the emergence of the structure of the giant connected components without considering tendrils. The reason is that breaking links transforms parts of the giant connected components into tendrils and, vice versa, adding links increases the giant connected components at the expense of tendrils and tubes.

In this Letter we reveal that an arbitrary directed graph with both unidirectional and bidirectional links has a rich hierarchical organization of layers of tendrils and tubes, see Fig. \ref{fig:picture}(b), that goes beyond the structure represented by the bow-tie diagram. We develop a computational algorithm that allows one to find all layers of tendrils and tubes. We also generalize the message-passing technique to directed graphs. This technique is used together with our algorithm to find the complete structure of directed networks.
We present the structures of some representative real-world networks and investigate how they are affected by random damage. We also introduce a generalized susceptibility and apply it to identify the percolation transition in the networks.

\textit{Structure of directed networks.}---In a directed graph $\mathcal{G}$, a giant strongly connected component ($G_S$) is a subgraph in which
any node can be reached from any other node by a directed path, always following links along their directions.
The set of nodes reachable from the $G_S$, following the directions of edges, is the giant out-component ($G_{out}$) and the set of nodes from which the $G_S$ is reachable, following the directions of edges, is the giant in-component ($G_{in}$), see Fig. \ref{fig:picture}(a) or \ref{fig:picture}(b).
These definitions give $G_S=G_{in} \cap G_{out}$ \cite{dorogovtsev2001giant}.
It is convenient to define the giant in-out component of a graph $\mathcal{G}$ as $G_{in} \cup G_{out}$. Note that in \cite{broder2000graph,newman2001random} $G_{in}$ and $G_{out}$ were defined without intersection.
If the directedness of edges is ignored, then  the corresponding giant connected component is called the giant weakly connected component ($G_W$) in the context of directed networks. There are also disconnected  finite clusters $F$ which,
together with $G_W$, form the entire graph $\mathcal{G} = G_W \cup F$.
$G_W$ includes the giant components, $G_{in}$, $G_{out}$, and, of course, $G_S$.
Moreover, $G_W$ includes finite directed components called tendrils ($T$).
Thus, $G_W=G_{in} \cup G_{out} \cup T$. Note that the finite components of $\mathcal{G}$ are $F \cup T$.

\textit{Tendril organization.}---Let us find how $T$ is organized.
First we find the set of nodes that are reachable from $G_{in}$ but are
not in $G_{in} \cup G_{out}$. We call this set the \textit{first-layer out-tendrils}, $T_{out}^{(1)}$ [see blue domains attached to $G_{in}$ in Fig. \ref{fig:picture}(b)].
Although this set is connected to $G_{in}$, it is more natural to call it out-tendrils, based on the direction of the
links by which they are connected to $G_{in}$.
Similarly, we find $T_{in}^{(1)}$,
the first-layer in-tendrils, which is the set of nodes from which $G_{out}$ can be reached, but are not in
$G_{in} \cup G_{out}$ [see blue domains attached to $G_{out}$ in Fig. \ref{fig:picture}(b)]. Tendrils $T_{in}^{(1)}$ and $T_{out}^{(1)}$ form the
first tendril layer, $T^{(1)} = T_{in}^{(1)} \cup T_{out}^{(1)}$ .
%
There is a special kind of tendrils which are simultaneously first-layer out-tendrils and first-layer in-tendrils [see dark blue domains connecting $G_{in}$ and $G_{out}$ in Fig. \ref{fig:picture}(b)]: we call these the first-layer tubes.

Let us introduce further tendril layers. The $n$th-layer
out-tendrils, $T_{out}^{(n)}$, is the set of nodes that are reachable from the tendrils $T_{in}^{(n-1)}$ in the $(n-1)$-th layer but do not belong to any previous layer [see green domains attached to the blue domains in Fig. \ref{fig:picture}(b)].
Similarly, the $n$th-layer in-tendrils, $T_{in}^{(n)}$, is the set of nodes from which $T_{out}^{(n-1)}$ can be reached but which do not belong to the previous layers.
The $n$th-layer tubes, $T_{tube}^{(n)}$, are tendrils that are simultaneously $n$th-layer out-tendrils and $n$th-layer in-tendrils, i.e.,
$T_{tube}^{(n)}=T_{in}^{(n)} \cap T_{out}^{(n)}$
[see the dark yellow domain connecting green domains in Fig. \ref{fig:picture}(b)].
The sets $T_{in}^{(n)}$, $T_{out}^{(n)}$ and $T_{tube}^{(n)}$ can be partitioned, respectively, into disjoint components -- individual in-tendrils, out-tendrils, and tubes-- which are shown as colored domains in Fig. \ref{fig:picture} (b). Individual tubes are the intersections of individual in-tendrils and out-tendrils.
In addition to the above sets of tubes, a single edge directed from any vertex $i$ in $G_{in} \setminus G_{S}$ to any vertex $j$ in $G_{out} \setminus G_{S}$, or from vertex $i$ in $T_{in}^{(n)}$ to vertex $j$ in $T_{out}^{(n)}$, is also a ``tube.'' Such edge-tubes (excluding the end vertices) must also be accounted for in the complete decomposition of a directed network.

\textit{Algorithm 1.}---We now present a computational algorithm for finding tendril layers in an arbitrary directed graph. Let $\hat{A}$ be the adjacency matrix of a directed graph: $A_{jk} = 1$ if there is an edge between nodes $j$ and $k$
in the direction of $k$, otherwise $A_{jk} = 0$. $\mathcal{N}_j$ is the set of neighbors of node $j$, irrespective of
direction. For any given graph the above method relies on first identifying the giant (largest) components: $G_S$, $G_{in}$, and
$G_{out}$. For this purpose one can use conventional cluster search algorithms which generally have time complexity $\mathcal{O}(N+L)$ for a graph of $N$ nodes and $L$ edges. Note that there may be multiple strongly connected components. In a finite graph, instead of a giant component, we consider the largest component.
We call the largest strongly connected component $G_S$ and its in- and out-components
$G_{in}$ and $G_{out}$.
Then we introduce a modified adjacency matrix,
$\hat{A}^{(1)}$, as follows: $A_{jk}^{(1)} = A_{kj}$
if $j,k \in (G_{in} \cup G_{out})$; $A_{jk}^{(1)} = A_{jk}$ if $j \notin (G_{in} \cup G_{out})$ for arbitrary $k$. In other words: we reverse the
direction of links where both end nodes are inside $G_{in} \cup G_{out}$ and leave the direction of all other links
unchanged. Using the modified adjacency matrix $\hat{A}^{(1)}$ we find the corresponding giant in- and out-components,
$G_{in}^{(1)}$ and $G_{out}^{(1)}$, in the modified graph.
Then we repeat the procedure for $\hat{A}^{(1)}$, reversing the direction of links inside $G_{in}^{(1)} \cup G_{out}^{(1)}$.
Thus we find $\hat{A}^{(2)}$ and the corresponding $G_{in}^{(2)}$ and $G_{out}^{(2)}$.
Repeating the same process we obtain a sequence of modified adjacency matrices and corresponding giant in- and out-components.
Using this process, we find the in-tendrils and out-tendrils in layer $n$,
\begin{eqnarray}
T_{in}^{(n)} = G_{in}^{(n)} \setminus (G_{in}^{(n-1)} \cup G_{out}^{(n-1)}),
  \label{eq:T_in_n} \\
T_{out}^{(n)} = G_{out}^{(n)} \setminus (G_{in}^{(n-1)} \cup G_{out}^{(n-1)}),
  \label{eq:T_out_n}
\end{eqnarray}
and tubes, $T_{tube}^{(n)}=T_{in}^{(n)} \cap T_{out}^{(n)}$.
Then each of these sets can be partitioned into
disjoint components (individual in-tendrils, out-tendrils, and tubes).
This algorithm also enables us to find all edge tubes in an arbitrary directed network \cite{[See Supplemental Material for the complete decomposition of a simple directed graph at ]timarSM}.


\textit{Algorithm 2.}---Diseases, injuries, and random or targeted damages impact the network structure described above. For a given realization of damage in a network, the impact can be found by applying Algorithm 1. In the case of random damage, a less time-consuming approximate algorithm can be found by generalizing the message-passing method in \cite{karrer2014percolation} to directed networks. Let a given graph be damaged by removing edges with probability $1-p$, in other words, any edge is present with probability $p$.
We introduce the probability generating function $H_{ij}^{(in)}(x)$ for the number of nodes reachable by
going from node $i$ to node $j$ against edge directions. Similarly, let
$H_{ij}^{(out)}(x)$ be the generating function for the number of nodes reachable by going from
node $i$ to node $j$
following the directions of the edges.
Assuming a large, locally treelike network, we can write
self-consistent equations,
\begin{eqnarray}
H_{ij}^{(in)}(x) = 1 + A_{ji} \Bigl[  -p + px \prod_{k \in \mathcal{N}_j \setminus i} H_{jk}^{(in)}(x)  \Bigr],
  \label{eq:H_self_cons1}  \\
H_{ij}^{(out)}(x) = 1 + A_{ij} \Bigl[  -p + px \prod_{k \in \mathcal{N}_j \setminus i} H_{jk}^{(out)}(x)  \Bigr].
  \label{eq:H_self_cons2}
\end{eqnarray}
Here $\mathcal{N}_j \setminus i$ is the set of neighbors of node $j$ excluding node $i$. Setting $x=1$ in Eqs. (\ref{eq:H_self_cons1}) and (\ref{eq:H_self_cons2}) we obtain a set of $4L$ coupled equations for the
$4L$ unknowns $H_{ij}^{(in)}(1)$ and $H_{ij}^{(out)}(1)$. These equations can be solved efficiently by iterations, using the
message-passing scheme
\cite{karrer2014percolation}.
Once a solution is found we
obtain the sizes of the giant in-component ($S_{in}$) and out-component ($S_{out}$),
\begin{eqnarray}
S_{in} = 1 - \frac{1}{N} \sum_{i=1}^N \prod_{j \in \mathcal{N}_i} H_{ij}^{(out)}(1),
  \label{eq:Si}  \\
S_{out} = 1 - \frac{1}{N} \sum_{i=1}^N \prod_{j \in \mathcal{N}_i} H_{ij}^{(in)}(1).
  \label{eq:So}
\end{eqnarray}
The size of the giant strongly connected component is
\begin{equation}
   \! S_S = \frac{1}{N} \sum_{i=1}^N  \Bigl[ 1{-}\prod_{j \in \mathcal{N}_i} H_{ij}^{(in)}(1) \Bigr] \Bigl[1{-}\prod_{j \in \mathcal{N}_i} H_{ij}^{(out)}(1) \Bigr].
  \label{eq:Ss}
\end{equation}
$S_{in}$, $S_{out}$, and $S_S$ depend on $p$ through Eqs. (\ref{eq:H_self_cons1}) and (\ref{eq:H_self_cons2}) that have a nontrivial solution if
$p > p_c$ and $p_c = 1 / \lambda^{(1)}$, where $\lambda^{(1)}$ is the largest eigenvalue  of the nonbacktracking matrix \cite{hashimoto2014automorphic}, as in ordinary
percolation \cite{karrer2014percolation}.

Algorithms 1 and 2 can be applied to networks consisting of both unidirectional and bidirectional links. Giant components of this kind of directed networks were first studied  by use of the generating function technique in \cite{boguna2005generalized}.
Also, setting $p=1$,
Algorithm 2 allows one to find the nodes belonging to the giant in- and out-components.

\begin{figure}[htpb!]
\centering
\includegraphics[width=8cm,angle=0.]{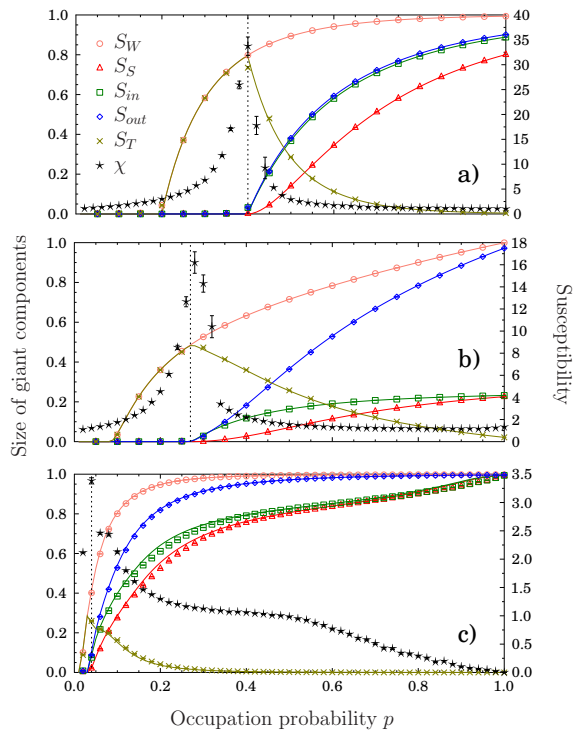}
\caption{Simulation (symbols) and message-passing results (solid lines) for the size of the giant weakly and strongly connected
components and the corresponding in- and out-components: (a) a directed Erd\H os-R\'enyi graph ($N = 10 000, \langle q_{tot} \rangle = 5$), (b) the
Gnutella P2P network ($N = 62 586, \langle q_{tot} \rangle = 4.726$ \cite{konect:2016:p2p-Gnutella31, konect:ripeanu02}), (c) the neural network of the male \emph{C. elegans}
($N = 495, \langle q_{tot} \rangle = 32.073$ \cite{jarrell2012connectome}). The dashed line corresponds to the critical parameter $p_c$ determined by
Eqs. (\ref{eq:H_self_cons1}) and (\ref{eq:H_self_cons2}) of the message-passing algorithm. The black stars and the right axis in panels (a)-(c) represent numerical simulations of the susceptibility $\chi$ from Eq. (\ref{eq:suscept}).
Simulation results correspond to averages over 100 realizations for (a),(b) and 1000 for (c).
Error bars represent the standard errors of the average values in this and all subsequent figures.}
\label{fig:giant}
\end{figure}

\textit{Structure of real networks}---Let us apply Algorithms 1 and 2 for studying the impact of random damage on directed complex networks.
In Fig. \ref{fig:giant} we present simulations of bond percolation (Algorithm 1) and the corresponding message-passing results (Algorithm 2) for three examples:
an Erd\H os-R\'enyi directed graph
(a random direction is assigned to each edge),
the Gnutella
peer-to-peer file sharing network from 2002
(\cite{konect:2016:p2p-Gnutella31, konect:ripeanu02}), and the neural network of \emph{Caenorhabditis elegans }(\emph{C. elegans}). The data
correspond to the neural network of the main body of the male \emph{C. elegans}, combining
both chemical and electrical synapses \cite{jarrell2012connectome}. We find that the message passing method gives a very good approximation in all cases.
Usually, message passing is expected to work well in large, sparse, treelike networks with low clustering coefficients, but it gives remarkably good
results even for a very small network with high clustering such as the \emph{C. elegans} neural network that consists of only 495
nodes and 7938 directed links and which has a clustering coefficient of $0.28$. This network also has many bidirectional links.

Analyzing these networks we observe an interesting asymmetry in the sizes of giant in- and out-components in the
damaged networks: the out-component is considerably larger than the in-component in both the Gnutella and the \emph{C. elegans} networks [see Figs. \ref{fig:giant} (b) and (c)].
In the undamaged \emph{C. elegans} network, the components $G_{in}$, $G_{out}$ and $G_S$ coincide, i.e., $G_{in}=G_{out}=G_S$, as is expected for a fully functional neural network. However, with increasing damage  these components become different and tendrils also appear. Moreover, the size $S_{in}$ of the giant in-component decreases considerably faster than the out-component. The cause of this asymmetry is related with the joint in-out degree distribution of nodes in the \emph{C. elegans} network.
Apart from the assortative correlation between in- and out-degrees, a striking feature of the distribution is that a large fraction of nodes (with varying in-degrees)
have exactly 2 outgoing links. The majority of these 77 nodes in the \emph{C. elegans} network are muscles: 65 body wall muscles and 5 male muscles.
Removal of the two outgoing connections destroys the feedback response of such a node and removes it from $G_{S}$, but it still belongs to $G_{out}$.
As a result, $G_{S}$ and $G_{in}$ decrease faster than $G_{out}$, in agreement with Fig. \ref{fig:giant} (c). This gives an example of how attacks on the
network structure can impact the functioning of the network due to the loss of feedback.
These findings may also imply an evolutionary compromise between ensuring the
tolerance of neural circuits to
random damage and minimizing the redundancy, which is the cost for the formation of multiple synaptic connections.

\begin{figure}[htpb!]
\centering
\includegraphics[width=\columnwidth,angle=0.]{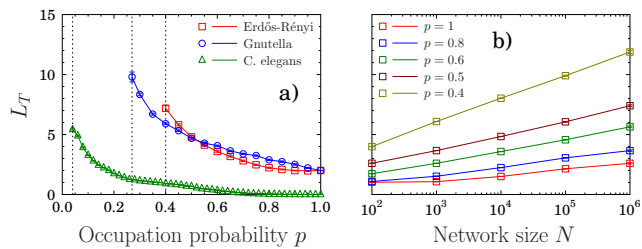}
\caption{(a) Mean number of tendril layers $L_T$ as a function of bond occupation probability $p$ for an Erd\H os-R\'enyi graph,
the Gnutella P2P network, and the neural network of the male \emph{C. elegans}. Dashed lines correspond to $p_c$ determined by the message-passing algorithm.
(b) $L_T$ versus network size $N$ for Erd\H os-R\'enyi networks of $\langle q_{tot} \rangle = 5$ at different values of $p$. All data points correspond to averages over at least 100 realizations of the network.}
\label{fig:tendril_nums}
\end{figure}

We use Algorithm 1 and find the number of tendril layers for a given realization of random damage in the networks in Fig. \ref{fig:giant}. Averaging over realizations gives the mean number of layers $L_T$ as a function of the bond occupation probability [see Fig. \ref{fig:tendril_nums}(a)].
Approaching the critical point $p_c$ from above $L_T$ increases monotonically for all
networks. Even for the small neural network of \emph{C. elegans}, $L_T$ exceeds 5 (on average) for high-enough damage.
Close to the critical point the definition of tendril layers becomes slightly dubious, as the typical size of tendril components
becomes comparable to that of the largest strongly connected component. Even in this case, however, our classification remains well defined
and meaningful.
In the Erd\H os-R\'enyi networks, we found that $L_T$ increases with increasing network size $N$ [see Fig. \ref{fig:tendril_nums}(b)].

\begin{figure}[htpb!]
\centering
\includegraphics[width=\columnwidth,angle=0.]{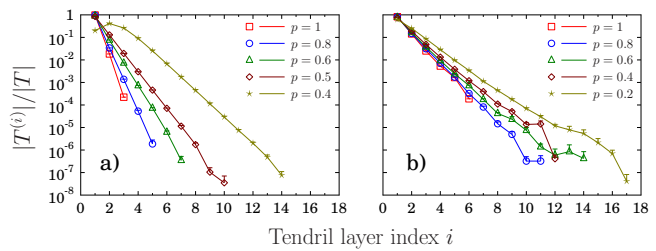}
\caption{Ratio of the size $|T^{(i)}|$ of the tendril layer $i$ to the total size $|T|$ of tendrils versus $i$ for different amounts of damage in (a) an Erd\H os-R\'enyi graph
($N = 10^6, \langle q_{tot} \rangle = 5$) and (b) the World Wide Web ($N = 875 713$, $\langle q_{tot} \rangle = 11.659$) \cite{konect:2016web-Google, konect:leskovec08}. Solid lines are for guidance only. The data correspond to averages over 100 realizations.
}
\label{fig:tendril_sizes}
\end{figure}

Using Algorithm 1, we find the size sequence of tendril layers
in a sample of the World Wide Web (obtained from Google \cite{konect:2016web-Google, konect:leskovec08}) and an Erd\H os-R\'enyi graph of similar size at different values of occupation probability. Figure \ref{fig:tendril_sizes} shows an approximately exponential decay for both networks,
with a decreasing rate of decay for increasing damage. This functional form of the size sequence of tendril layers is valid even near the critical point, where the rate of decay
approaches a certain value.

\textit{Susceptibility.}---
In order to quantitatively characterize the response of a directed network to damage, we introduce a generalized susceptibility,
\begin{equation}
  \chi=\frac{1}{N}\sum_{i,j\in \mathcal{G} \backslash (G_{in} \cup G_{out})}C(i,j),
  \label{eq:suscept}
\end{equation}
where indices $i$ and $j$ run only over nodes belonging to the finite components (tendrils and disconnected finite clusters).
Nodes in the giant in-out component (the order parameter) are excluded. The correlation function $C(i,j)$ is defined as follows: (i) $C(i,i)=1$ and (ii) $C(i,j)=1$ if there is a directed path from $i$ to $j$ either along or against the edge directions, or in both directions. Otherwise, $C(i,j)=0$. Note that this definition of $\chi$ is valid for any directed graph $\mathcal{G}$ with clustering, degree correlations, bidirectional edges, and so on.
Equation (\ref{eq:suscept}) generalizes the susceptibility of the one-state Potts model \cite{kasteleyn1969phase,wu1982potts} to the case of directed networks.
In this context $\chi$ has the meaning of the mean number of nodes in the finite components ($F \cup T$)  reachable from a randomly chosen node also in the finite components,
following edges either along or against the edge directions.
The divergence of $\chi$ signals the percolation transition at
$p=p_c$ in the limit $N \rightarrow \infty$. At a finite $N$, $\chi$ has a maximum. Results of simulations showing this behavior are displayed in Fig. \ref{fig:giant} for
the Erd\H os-R\'enyi, Gnutella, and \emph{C. elegans} networks. The position of the maximum agrees very well with $p_c$ predicted by the message-passing algorithm.
When a directed network approaches $p_c$ from the ordered state, we have $G_S \rightarrow 0$; this is in contrast with the giant weakly connected component $G_W$, which remains nonzero. The distribution of finite clusters is also not changed qualitatively around $p_c$. Thus the main contribution to the divergence of $\chi$ is given by nodes in tendrils which show critical statistics.
Analyzing the susceptibility near $p_c$ in the Erd\H os-R\'enyi network in Fig. \ref{fig:giant}(a), we find the standard mean-field behavior $\chi \sim |p-p_c|^{-1}$ both above and below $p_c$. If the edge directedness is neglected, then Eq. (\ref{eq:suscept}) determines the susceptibility for ordinary percolation \cite{kasteleyn1969phase}. In this case, indices $i$ and $j$ run over nodes belonging to disconnected finite clusters.
An analytical consideration of $\chi$ will be given elsewhere.


In conclusion, we have developed algorithms enabling us to find the entire
structure of an arbitrary directed network. We focused on
tendrils and tubes, which were shown in the original
bow-tie diagram \cite{broder2000graph}, but until now had not attracted serious
attention. We revealed that the array of tendrils
and tubes in a directed network actually has a rich hierarchical,
layered architecture. We found that random damage increases the
number and size of tendril layers, decreases the sizes of giant in-
and out-connected components, and enhances the susceptibility of
directed networks to damage. The tendril layers and giant
components are closely interrelated, and we suggest that our
concept of the hierarchical organization of directed networks
and our algorithms will be useful for understanding functions
of real networks of this class and their tolerance to failures and
attacks.

\emph{Acknowledgements.}--This work was supported by the FET proactive IP project MULTIPLEX 317532 and Grant No. PEST UID/CTM/50025/2013.


\bibliography{bibliography_directed}

\end{document}